\documentclass[conference]{IEEEtran}
\IEEEoverridecommandlockouts
\usepackage{cite}
\usepackage{amsmath,amssymb,amsfonts}
\usepackage{algorithmic}
\usepackage{graphicx}
\usepackage{textcomp}
\usepackage{xcolor}
\usepackage{multirow}
\usepackage{float}
\def\BibTeX{{\rm B\kern-.05em{\sc i\kern-.025em b}\kern-.08em
    T\kern-.1667em\lower.7ex\hbox{E}\kern-.125emX}}
\begin{document}

\title{PICNIC: Silicon Photonic Interconnected Chiplets with Computational Network and In-memory Computing for LLM Inference Acceleration\\
\thanks{This work is funded in part by the National University of Singapore through the Microelectronics Seed Grant (FY2024); and in part by the National Research Foundation (NRF), Singapore, under the Competitive Research Programme (Award NRF-CRP24-2020-0002 and NRF-CRP24-2020-0003)}
\thanks{Corresponding author: Xuanyao Fong}
\thanks{\textsuperscript{\dag} Both authors contributed equally to this work.}
}

\author{\IEEEauthorblockN{Yue Jiet Chong$^1$\textsuperscript{\dag}, Yimin Wang$^2$\textsuperscript{\dag}, Zhen Wu$^3$, Xuanyao Fong$^4$}
\IEEEauthorblockA{\textit{Department of Electrical and Computer Engineering} \\
\textit{National University of Singapore}, Singapore 
\\
Email: \{jason.yj.chong$^1$, kelvin.xy.fong$^4$\}@nus.edu.sg, \{yimin.wang$^2$, e0323083$^3$\}@u.nus.edu}
}

\maketitle

\begin{abstract}
This paper presents a 3D-stacked chiplets based large language model (LLM) inference accelerator, consisting of non-volatile in-memory-computing processing elements (PEs) and Inter-PE Computational Network (IPCN), interconnected via silicon photonic to effectively address the communication bottlenecks.
A LLM mapping scheme was developed to optimize hardware scheduling and workload mapping. Simulation results show it achieves $3.95\times$ speedup and $30\times$ efficiency improvement over the Nvidia A100 before chiplet clustering and power gating scheme (CCPG).
Additionally, the system achieves further scalability and efficiency improvement with the implementation of CCPG to accommodate larger models, attaining $57\times$ efficiency improvement over Nvidia H100 at similar throughput.
\end{abstract}


\section{Introduction}

Large language model (LLM) inference involves static and dynamic data.
The static data are pre-trained weights obtained during supervised training of the model, whereas the dynamic data are temporary data generated during query-key-value (QKV) projections in the attention layers \cite{llm_data_movement}.
In most existing LLM accelerator architectures, both static and dynamic data are constantly moved between main memory (RAM) and computing units \cite{llm_memory_bottleneck}.
Recent trends in the development of LLMs show a dramatic increase in model size \cite{LLM_trend}, implying that the volume of data transfer between memory and compute units will increase and lead to increasing power consumption and processing latency \cite{llm_power_demand}.

\begin{figure}[t]
    \centering
    \includegraphics[width=0.9\linewidth]{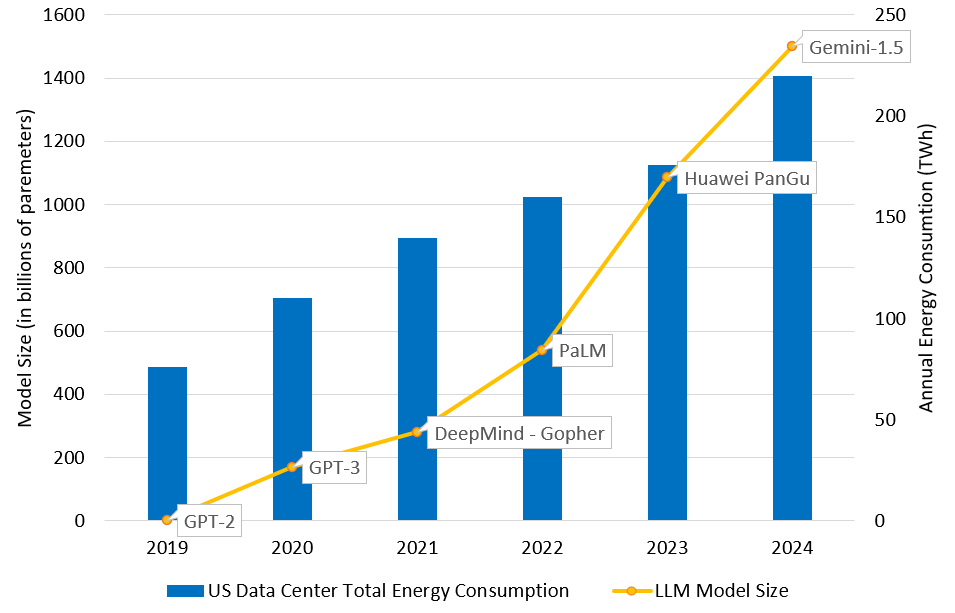}
    \caption{LLM Model Size and US Data Center Total Energy Consumption}
    \label{fig:llm_trend_energy}
\end{figure}

In response, researchers are exploring alternative hardware architecture design approaches \cite{data_mapping, nonVN_survey} to address the increasing volume of data movement. Nevertheless, new challenges arise due to differences in hardware scheduling, data locality, and workload mapping \cite{mapping_nonVN_arch, iccad_leap}. These challenges must be addressed to fully exploit hardware resources while maintaining scalability, energy efficiency, and improving system performance.

Chiplet-based system design has emerged as a promising approach to enhance scalability in modern VLSI architectures \cite{chiplet_survey}. However, this implementation introduces challenges in energy efficiency, particularly in electrical chip-to-chip (C2C) communication. In fact, communication can account for as much as 30\% of the total energy consumption in VLSI systems \cite{commun_energy}. For instance, electrical C2C communication typically incurs an energy cost of 3 pJ/bit, while off-chip memory access consumes up to 30 pJ/bit \cite{si_photonic_survey}. These energy costs become especially significant in highly parallel multi-core systems, such as the GPUs and NPUs used for AI workloads—where massive data movement leads to communication energy consumption that is comparable to computational energy \cite{us_datacenter_energy}.

To mitigate the LLM inference bottlenecks, we studied and implemented hardware-software co-design for LLM inference accelerator, named PICNIC, with the following characteristics:

\begin{itemize}
    
    \item An Inter-PE Computational Network (IPCN) as interconnecting platform for In-memory Computing (IMC) macros, integrated with computational capabilities and re-programmability via a dedicated instruction set for efficient data flow.
    
    \item Heterogeneous 3D Stacked-IC compute tile to increase computing, communication, and area efficiencies via vertical integration of chiplets in different domains, \textit{i.e.}, digital, analog, and optical.

    \item Temporal hardware scheduling that incorporates dedicated context window tiling and efficient key-value caching (KV cache), ensuring balanced network traffics and utilization of processing elements (PEs).

    \item Chiplet clustering scheme with power gating technique (CCPG) ensuring system power scales sub-linearly for increasing model sizes.  

\end{itemize}

The simulation results of PICNIC show 3.95$\times$ speedup and 30$\times$ efficiency improvement over Nvidia A100 in Llama-8B inference, as well as 57$\times$ efficiency improvement over Nvidia H100 at similar throughput via CCPG.

\begin{figure}[t]
    \centering
    \includegraphics[width=1\linewidth]{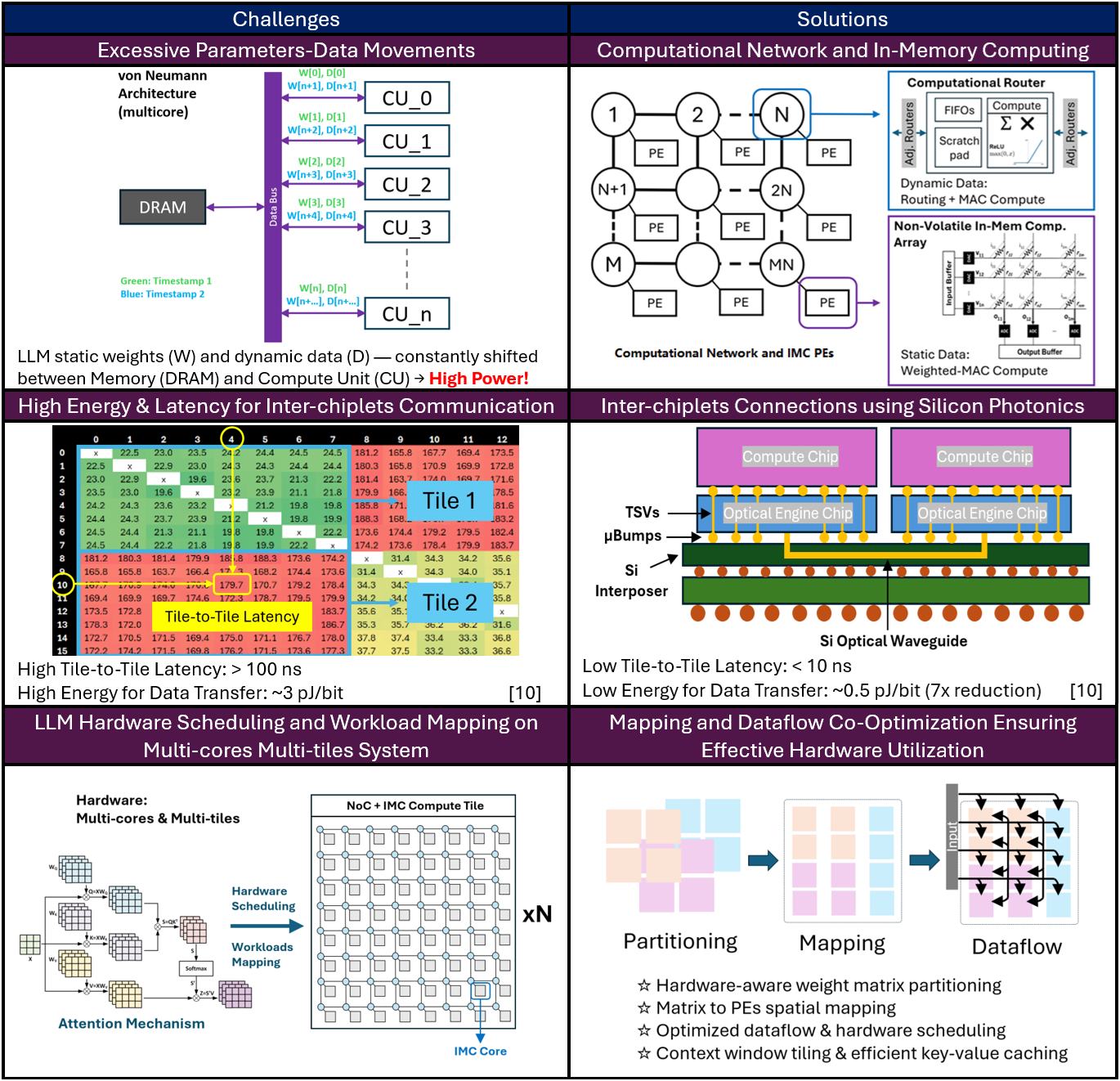}
    \caption{Challenges and Solutions for LLM Inference}
    \label{fig:prelim_1}
\end{figure}

\section{PICNIC Hardware Architecture}
\begin{figure*}[t]
    \centering
    \includegraphics[width=1\linewidth]{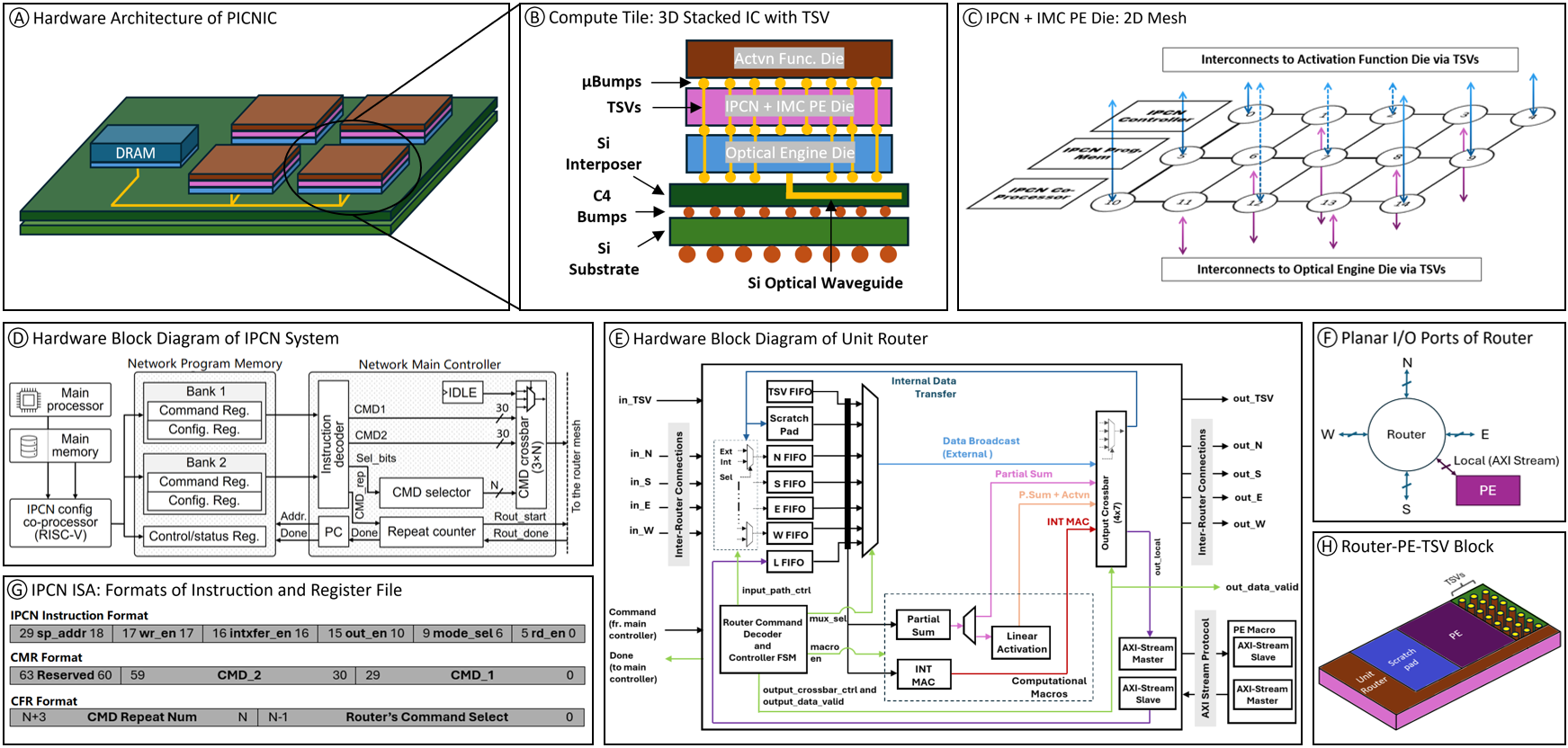}
    \caption{PICNIC Hardware Architecture: IPCN, Computing Macros, Interconnects and Instruction Set}
    \label{fig:method}
\end{figure*}

The hardware architecture of PICNIC LLM Inference Accelerator is shown in Fig. \ref{fig:method}(a). It consists of multiple compute tiles (CT), each of which is implemented as a 3D-Stacked IC (3D-SIC) for heterogeneous design shown in Fig. \ref{fig:method}(b). These CTs are interconnected with silicon photonics for inter-tile data transfer and memory access (DRAM). The DRAM acts as a hub for external data communication. Each CT consists of multiple processing elements (PEs) interconnected via the 2D-mesh Inter-PE Computational Network (IPCN). The PEs perform static weight multiply-accumulate (SMAC) operations while the IPCN conducts dataflow control and dynamic data multiply-accumulate (DMAC) operations.

\subsection{Processing Element (PE)}
The PE consists of a non-volatile resistive random access memory array compute-in-memory macro (RRAM-CIM) \cite{rram_imc} to perform SMAC operations. Each unit of RRAM cell stores a unit weight/parameter of the neural networks as the resistance state. Due to the non-volatile nature of RRAM, the weights only need to be programmed once for a given model, significantly reducing reconfiguration overhead. Once initialized, the PEs perform SMAC operations directly in the analog domain, leveraging the inherent parallelism and energy efficiency of in-memory computing.

To mitigate hardware non-idealities, the macro incorporates a feedback-loop calibration mechanism. During the initialization phase, it calibrates the macro to fully utilize the ADC input swing, thereby minimizing discretization errors. Offsets identified during calibration are stored for subsequent compensation during inference. Furthermore, the macro employs voltage-mode sensing, which inherently normalizes the wide variations in output dynamic range \cite{rram_imc}.

\subsection{Inter-PE Computational Network (IPCN)}
The IPCN is designed as an interconnecting platform for In-memory Computing (IMC) macros to orchestrate dataflow and perform various computations on the network data to accommodate different AI workloads, as shown in Fig. \ref{fig:method}(c). For computations, SMAC operation is performed in PE while other mathematical operations, \textit{e.g.} DMAC, are performed in the routers. The IPCN consists of three main parts, \textit{i.e.}, Program Memory, Network Main Controller (NMC), and 2D-mesh of routers with PE, shown in Fig. \ref{fig:method}(d). It supports re-programmability via a dedicated instruction set for efficient control of data flow.

\subsubsection{Network Program Memory (NPM)}
The NPM stores the instructions that control the operations of routers and processing elements. The NPM consists of 3 Banks, \textit{i.e.}, Bank 1 (B1), Bank 2 (B2), and Control/Status Register Bank (CSR). Each of B1 and B2 consists of two sub-banks: the command register (CMR) and the configuration register (CFR). In each row, the CMRs store 2 different commands to be executed by the routers while the CFRs store the command selection signal for each router and the number of command repetitions. In parallel, each router will combine information from the CMR and CFR to determine whether to IDLE or to execute CMD1 or CMD2, and to repeat the same operation by the number of times defined in CFR. 

\subsubsection{Configuration Co-processor}
The NPM is configured alternately between B1 and B2 by the co-processor, using firmware stored in the system main memory.
Concurrently, the Network Main Controller (NMC) performs sequential reads from the NPM.
Specifically, while the NMC is reading data from B2, the co-processor configures B1, and vice versa.
This interleaved configuration and access mechanism minimizes IPCN idle cycles during runtime.
Also, this approach ensures continuous data flow and enhances overall system throughput.

\subsubsection{Network Main Controller (NMC)}
The NMC reads and decodes the contents in NPM to control the operations of routers and PEs within the 2D-mesh network and establish a dedicated data flow. The sub-modules are as follows: (i) \textit{Instruction Decoder}: Decodes the instruction into three sections, \textit{i.e.}, routing command, command selection and number of command repetitions. (ii) \textit{Command Crossbar}: A 3-input-N-output crossbar (N: number of routers in the network). Each individual router is fetched with either CMD1, CMD2 or IDLE based on their selection signals from the Command Selector. (iii) \textit{Command Repeat Counter}: Stores the number of repetitions for the command and decrements by 1 when each command completes execution.

\subsubsection{Unit Router}
The unit router in the IPCN 2D Mesh, as shown in Fig. \ref{fig:method}(e), has two main functions: data packets routing and in-network computing. The sub-modules are as follows:
(i) \textit{Data I/O ports}: 4 planar ports for inter-router connections, a pair of AXI-Stream adapters for router-PE connection (shown in Fig. \ref{fig:method}(f)) and 2 vertical ports with through-silicon-via (TSV). Each port is integrated with First-In, First-Out buffer (FIFO) for temporary data storage. (ii) \textit{Decoder and Controller}: Decodes the command from NMC and controls the operations of each macro in the unit router. (iii) \textit{Computational Macros}: Enables digital in-network computing on data stored in the router, optimized for AI workload. The macros include partial summation, linear activation and DMAC. Each unit router is attached with a PE, forming a router-PE pair.

\subsubsection{IPCN Instruction Set Architecture (ISA)}
The IPCN instruction is a 30-bit vector, as shown in Fig. \ref{fig:method}(g), consisting of the following sub-fields: (i) \textit{rd\textunderscore en} indicates FIFO indices for data read, (ii) \textit{mode\textunderscore sel} to select the operation mode of the router, (iii) \textit{out\textunderscore en} indicates output directions of data packet, (iv) \textit{intxfer\textunderscore en} for internal data movement between FIFOs and scratchpad memory within the router, and (v) \textit{SP\textunderscore addr} as address to access the scratchpad memory. For data movement in the IPCN, unicast and broadcast are supported. Unicast moves data in one direction whereas broadcast moves data in multi-directions (up to all I/O ports).

A toolchain consists of an application programming interface (API) and a program compiler is developed in Python to facilitate the hardware utilization. The API is a library containing the ISA, enabling the user to develop firmware for system data flow control based on the AI workload. The compiler converts the user program into a hex file to be loaded into the NPM.

\begin{figure}[t]
    \centering
    \includegraphics[width=0.9\linewidth]{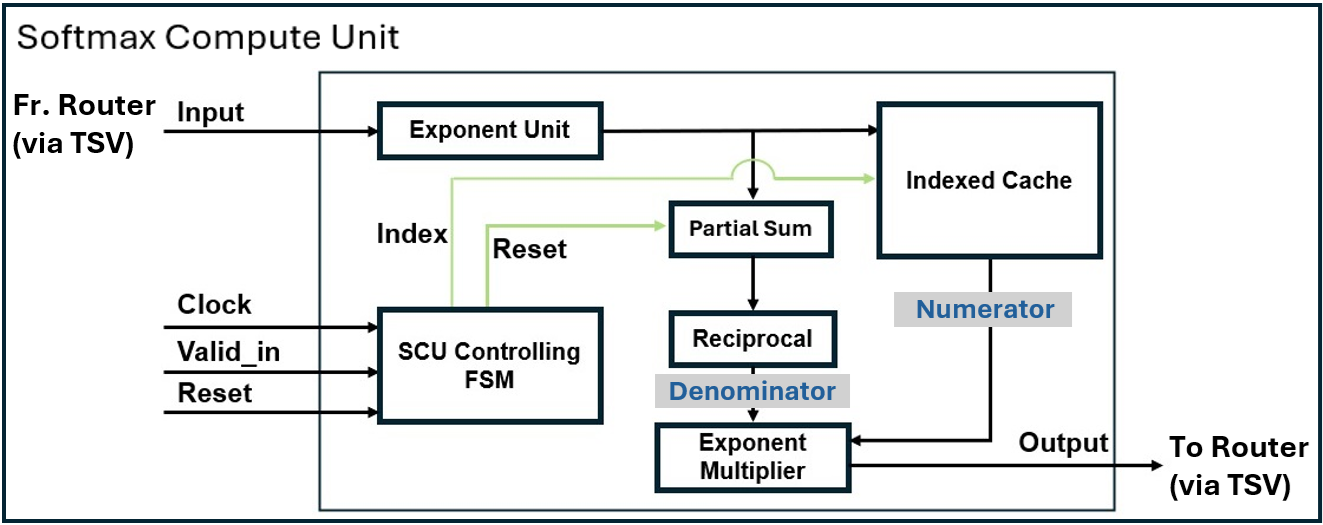}
    \caption{Softmax Compute Unit}
    \label{fig:scu}
\end{figure}

\subsection{Softmax Compute Unit (SCU)}

\text The SCU, shown in Fig. \ref{fig:scu},  is controlled by a finite state machine (FSM) that has three different states.
Initially, the SCU receives inputs from the router sequentially.
The calculated exponential results is sent to both indexed cache and partial sum adder.
In the second state, when the full input data sequence has been transmitted, the partial sum result is used to calculate its reciprocal as the denominator of the softmax function.
In the last state, the multiplier unit multiplies the reciprocal with the data in the cache (nominator) to produce the final softmax result. 
The SCU then switches between state~2 and state~3 to produce a continuous output.
The exponential function is expressed with an eight-segment piece-wise linear approximation. 

\subsection{3D-Stacked IC and Optical Inter-chiplets Connection}
3D-SIC consists of heterogeneous dies that are vertically stacked and interconnected using TSVs \cite{3d_sic_tsv}.
As shown in Fig.~\ref{fig:method}(b), the top chiplet consists of the activation function macros, which perform computations of non-linear activation functions such as softmax (SCU).
The next highest die consists of the IPCN 2D-Mesh and PEs for dataflow orchestration, SMAC, and DMAC.
The bottom die is the optical engine to enable efficient C2C communications utilizing silicon photonics \cite{si_photonics_hardware_analysis}.
It consists of an optical transceiver including a laser source, waveguides, microring modulators (MRM), network switching elements, and photo-detectors.
The silicon optical waveguide is embedded in the silicon substrate \cite{si_photonics_integration}, forming an optical network connecting all the chiplets of the system.

The TSVs are allocated in an alternating column-wise pattern within the IPCN, \textit{i.e.}, TSVs in odd-numbered columns connect to the top die, whereas those in even-numbered columns connect to the bottom die, as shown in Fig.~\ref{fig:method}(c). This arrangement reduces the TSV density, thereby mitigating signal interference and improving inter-die communication reliability \cite{tsv_crosstalk}.

\begin{figure}[t]
    \centering
    \includegraphics[width=0.7\linewidth]{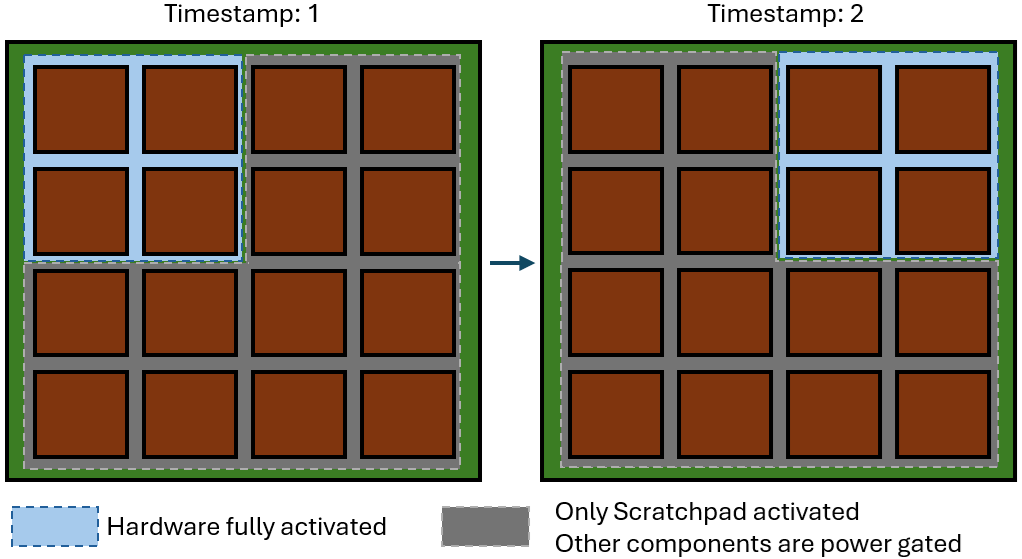}
    \caption{Illustration of Chiplet Clustering with Power Gating Scheme}
    \label{fig:chip_cluster}
\end{figure}

\subsection{Chiplet Clustering and Power Gating (CCPG)}
For running an LLM, the workloads are executed in a sequential, layer-by-layer manner. During the computation of a given layer, all other layers remain idle, which presents an opportunity for power optimization. To exploit this characteristic, PICNIC adopts a chiplet-based layer-wise weights allocation strategy, wherein each layer is mapped to a cluster of adjacent chiplets (details are discussed in \textit{Section III}). For CCPG, as shown in Fig. \ref{fig:chip_cluster}, four adjacent compute-tile chiplets are grouped as a cluster. During computation runtime, only one cluster is fully activated; whereas for all other clusters, only the scratchpad memory modules stay activated for context window data retention (KV caching) while other hardware macros are power gated (sleep mode). This selective activation significantly reduces system power consumption. The weights stored in RRAM are unaffected due to its non-volatility. 

\section{PICNIC LLM Inference Orchestration}

    The chiplets handle the LLMs in a layer-wise manner: each chiplet stores an attention layer or a feed-forward layer. 
    For example, Llama 3.2-1B holds 16 decoders, where each decoder comprises an attention layer and three feed-forward layers. 
    An end-to-end partitioning, mapping, and scheduling scheme orchestrates the processing of each layer, ensuring balanced network traffic and utilization of PEs. 

    \subsubsection{Partitioning}
    In each layer, both static and dynamic data are partitioned to meet the capacity limit of the PE crossbar arrays and local scratchpads. 
    Partitioning is applied along both row and column dimensions of the matrices. 
    Partitioning static weight matrices, $\mathbf{W_Q}$, $\mathbf{W_K}$, $\mathbf{W_V}$, and $\mathbf{W_O}$ $\in \mathbb{R}^{D\times D}$ incurs extra collective communications for partitioned input broadcast and partial output reduction along the embedding dimensions $D$. 
    Partitioning the intermediate data, $\mathbf{Q}$, $\mathbf{K}$, $\mathbf{V}$ $\in \mathbb{R}^{S\times D}$, and $\mathbf{S}$ $\in \mathbb{R}^{S\times S}$, involves both embedding dimensions and the sequence length $S$, which relates its temporal scheduling to the attention mechanism, \textit{e.g.}, softmax activation and KV cache. 

    \subsubsection{Mapping}
\begin{figure}[t]
    \centering
    \includegraphics[width=0.85\linewidth]{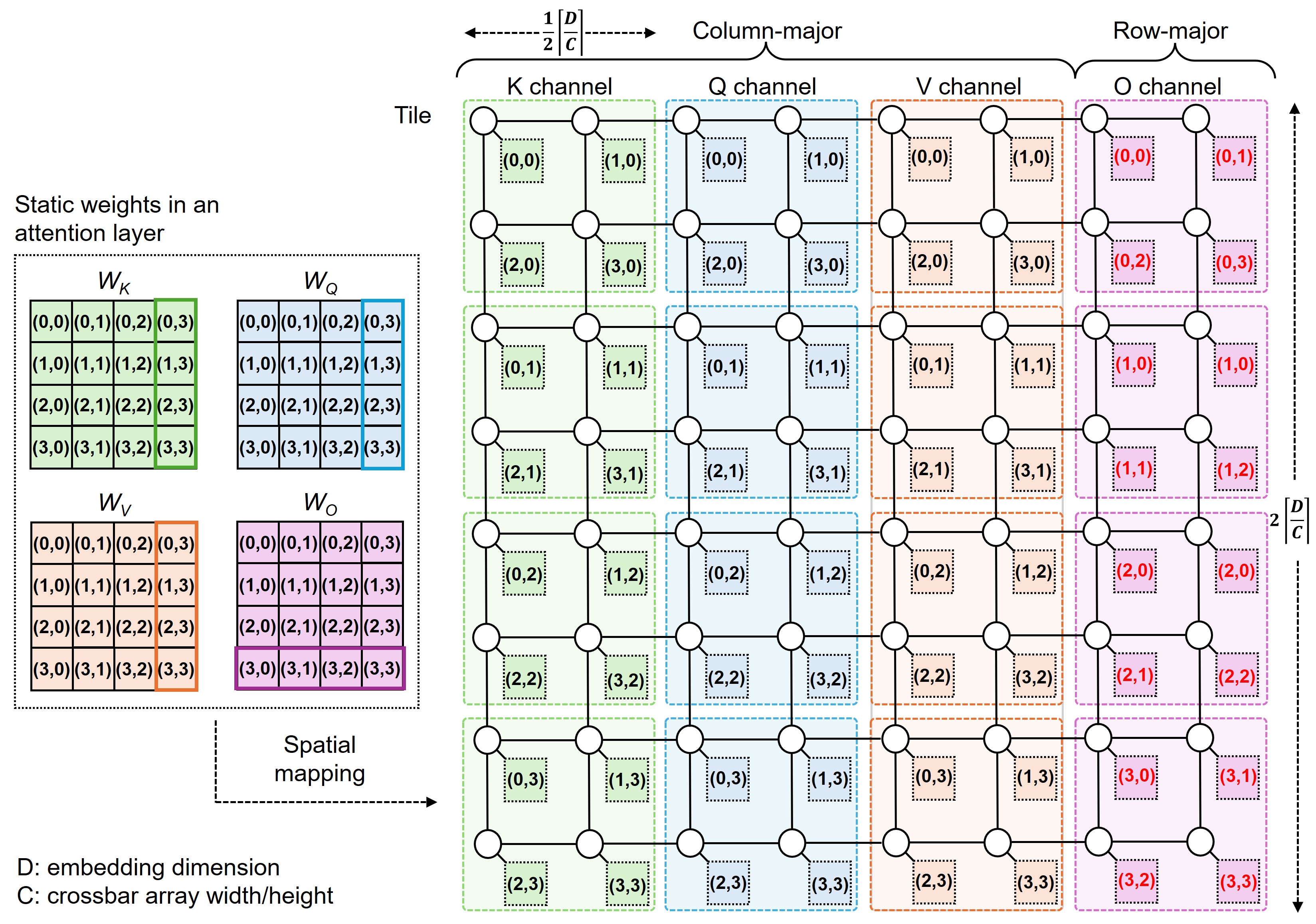}
    \caption{Spatial mapping of the weight matrices in an attention layer within a chiplet of PICNIC (with respect to K-Q-V-O Channels)}
    \label{fig:mapping}
    \end{figure}

    The partitioned $\mathbf{W_Q}$/$\mathbf{W_K}$/$\mathbf{W_V}$/$\mathbf{W_O}$ are spatially mapped to the PE crossbar arrays and the partitioned $\mathbf{Q}$/$\mathbf{K}$/$\mathbf{V}$/$\mathbf{S}$ are mapped into the distributed scratchpad. 
    For mapping on PE crossbar arrays, each matrix is heuristically constrained which to be mapped in a column-wise rectangular region and optimize the mapping by tweaking three factors: intra-matrix shape, inter-matrix shape, and row-column order. 
    The optimized mapping scheme adopted in PICNIC is illustrated in Fig.~\ref{fig:mapping}. 
    For mapping on scratchpads, the intermediate matrix is stored in the scratchpads within the region of its weight matrix, \textit{i.e.}, $\mathbf{Q}$ is stored in the scratchpads of the router-PE pairs where $\mathbf{W_Q}$ has been pre-placed, which enables output reduction in the vicinity. 

    \subsubsection{Scheduling}
    Scheduling handles the temporal dataflow across the PE array and IPCN. 

    \textbf{FlashAttention}: A kernel-fused attention mechanism, FlashAttention~\cite{NeurIPS-2022-FlashAttention} is adopted in this work.
    FlashAttention spawns a two-level nested loop computing flow. The inner loop is partially unrolled and executed in parallel to fully utilize the DMAC resources in IPCN. 

    \textbf{KV cache}: The \textbf{K/V} vectors corresponding to the tokens generated in the decode phase are appended to the scratchpads pre-allocated to $\mathbf{K}$/$\mathbf{V}$. 
    The $\mathbf{K}$/$\mathbf{V}$ vectors are cyclically stored in the different pre-allocated scratchpads, which enables a balanced utilization of the distributed scratchpads regardless of the length of the sequence being processed. 

    \textbf{Collective communication}: The reduction and broadcast are determined by the spanning tree algorithm, where the data traffic is balanced and non-congestive due to the regular and aligned mapping. 
\section{System Modeling and Evaluation}
\begin{figure}[t]
    \centering
    \includegraphics[width=0.75\linewidth]{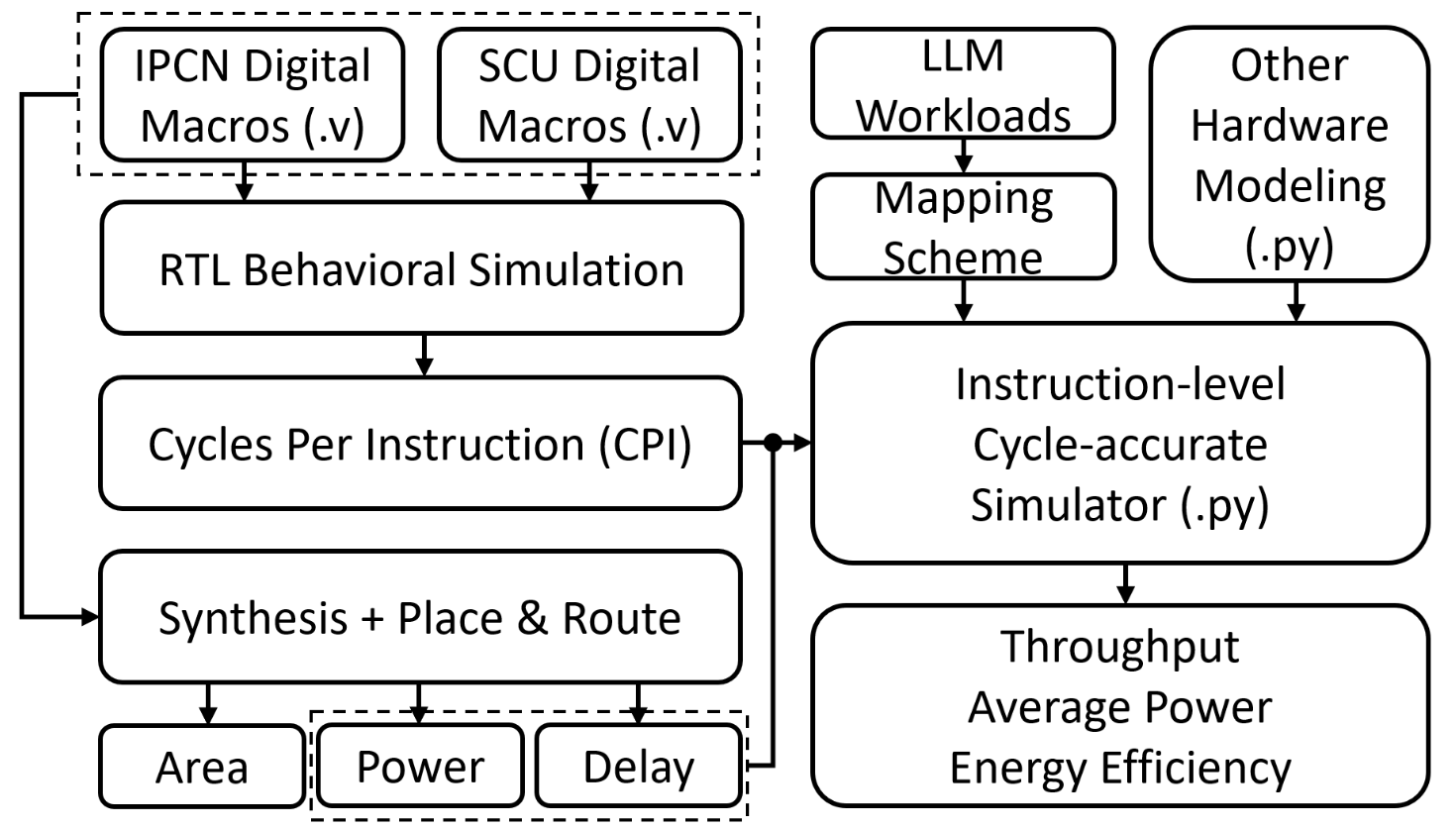}
    \caption{Overview of System Modeling and Evaluation}
    \label{fig:eval_flow}
\end{figure}

The PICNIC LLM inference accelerator is evaluated via hardware-software co-verification as shown in Fig.~\ref{fig:eval_flow}. 
The digital hardware of the system was developed and verified using Verilog HDL.
Hardware Synthesis and Place \& Route (P\&R) were performed using \textit{Synopsys Design Compiler} and \textit{Cadence Innovus} respectively.
The power and area of the scratchpad memory macro are obtained using \textit{CACTI} \cite{CACTI}.
Other hardware blocks are modeled and emulated in software using mathematical models.
For LLM workload mapping and hardware scheduling, a mapping scheme with fine-grained model partitioning, heuristically optimized spatial mapping and temporal scheduling was developed.
Inference emulation and benchmarking are performed using an instruction-level cycle-accurate simulator via the IPCN API. 

RRAM non-idealities are addressed through a combination of software technique, such as noise-resilient neural network training for conductance relaxation \cite{rram_imc}, and hardware solutions described in \textit{Section II-A}. Consequently, this aspect is not the primary focus of this work.

\begin{table}[t]
    \caption{PICNIC System Parameter}
    \centering
    \setlength{\tabcolsep}{9pt}
    \begin{tabular}{c|c|c|c}
         \hline
         \multicolumn{4}{c}{\textbf{System Level}} \\
         \hline
         Bit-width & 64 & Frequency & 1 GHz \\
         \hline \hline
         \multicolumn{4}{c}{\textbf{Tile Level}} \\
         \hline
         IPCN Dimension & 32$\times$32 & Softmax CU \# & 1024 \\
         \hline \hline
         \multicolumn{4}{c}{\textbf{Macro Level (per unit Router-PE pair)}} \\
         \hline
         PE Array Size & 256$\times$256 & non-weighted MAC \# & 16 \\ 
         \hline
         Scratchpad Size & 32 KB & I/O Ports \# & 7 \\ 
         \hline
         FIFO Size (each) & 256 B & TSV Dimension & 32$\times$2 \\ 
         \hline
    \end{tabular}    
    \label{tab:sys_param}
\end{table}

\subsection{Performance Evaluation}
The performance of PICNIC is evaluated on LLMs as shown in Table \ref{tab:benchmark}. The average system power increases with the model size because more chiplets are activated to accommodate the model weights. Concurrently, the throughput of the system reduces as more data movement and computations occur within the 2D-mesh, leading to higher overall latency. Thus, the energy efficiency (expressed in tokens/J) decreases exponentially.

For various context lengths using the same model, the average power consumed by the system reduces slightly with increasing context length. This is due to the reduction in rate of C2C communications, which is discussed in \textit{Section~IV-C}.

Table \ref{tab:cross_platform} shows the performance and energy efficiency comparisons of PICNIC to other platforms with various architectures. The evaluations are based on Llama-8B (1024/1024, batch size 1) with Nvidia H100 as baseline. PICNIC achieves superior energy efficiency as it eliminates weights transfer, along with minimal dynamic data transfer between compute unit and main memory during inference. The former is due to weights storage and SMAC operations in non-volatile IMC PE and the latter with IPCN coupled with efficient hardware scheduling as well as the efficient KV caching via scratchpad memory next to the PE. In contrast, GPUs like A100 and H100 have high utilization of ALU-main memory communications for both dynamic data and weights, which incur high power and latency.

\begin{table}[t]
    \caption{Benchmark of LLM Inference for PICNIC}
    \centering
    \setlength{\tabcolsep}{3.8pt}
    \begin{tabular}{c|c|c|c|c}
         \hline
         \multirow{2}{*}{Model}
         & Context Length & Throughput & Average & Efficiency\\
         & (Input/Output) & (tokens/s) & Power (W) & (tokens/J)\\ 
         \hline
         \multirow{3}{*}{Llama 3.2 – 1B*}
         & 512/512 & 1503.8 & 4.0520 & 371.1 \\
         & 1024/1024 & 969.2 & 4.0513 & 239.2 \\
         & 2048/2048 & 566.4 & 4.0507 & 139.8 \\
         \hline
         \multirow{3}{*}{Llama 3 – 8B*}
         & 512/512 & 386.5 & 28.4018 & 13.6 \\
         & 1024/1024 & 309.8 & 28.4015 & 10.9 \\
         & 2048/2048 & 221.9 & 28.4010 & 7.8 \\
         \hline
         \multirow{3}{*}{Llama 2 – 13B*}
         & 512/512 & 228.9 & 52.3014 & 4.4 \\
         & 1024/1024 & 192.4 & 52.3012 & 3.7 \\
         & 2048/2048 & 146.2 & 52.3009 & 2.8 \\
         \hline
    \end{tabular}
    
    \vspace{0.15cm}
    {\raggedright *Power and Efficiency without chiplet clustering and power-gating \\}
    \label{tab:benchmark}
\end{table}

\begin{figure}[t]
    \centering
    \includegraphics[width=1\linewidth]{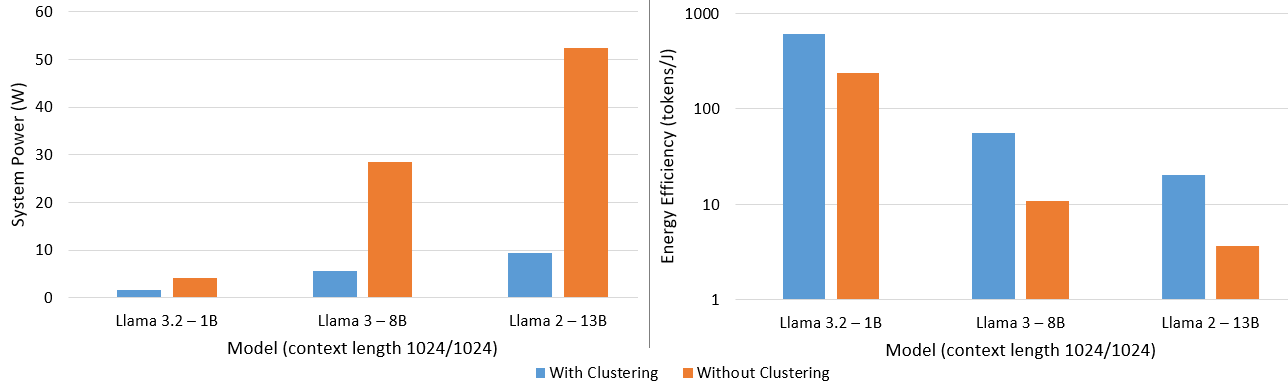}
    \caption{Comparisons of System Power and Energy Efficiency with and without Chiplet Clustering and Power Gating Scheme}
    \label{fig:cluster_compare}
\end{figure}

\begin{table*}[t]
    \caption{Comparison with Other Platforms}
    \centering
    \setlength{\tabcolsep}{3.8pt}
    \begin{tabular}{c|c|c|c|c|c|c|c}
         \hline
         Platform & This Work & TransPIM \cite{transpim} & Cambricon-LLM \cite{cambricon_llm} & NV A100 & NV H100 & Apple M4-Max & Cerebras-2 \cite{cerebras_wse} \\
         \hline
         \multirow{2}{*}{Architecture} 
         & SiPh Chiplets based & Hybrid PIM-NMC & Chiplets based NPU- & \multicolumn{2}{c|}{Multi-core} & \multirow{2}{*}{SoC-NPU} & Wafer-Scale \\
         & IPCN \& A-IMC & in HBM & NAND Flash PIM & \multicolumn{2}{c|}{GPU} &  & Engine \\
         \hline
         Throughput* (tokens/s) & 309.83 & 270 & 36.34 & 78.36 & 274.26 & 69.77 & 1800 \\
         \hline 
         Average Power (W) & $^{\dag}$5.6 & 40 & 36.3 & 200 & 280 & 80 & 15000\\
         \hline
         Energy efficiency (tokens/J) & 55.38 & 6.8 & 1 & 0.39 & 0.98 & 0.87 & 0.12 \\
         \hline
         Speedup\^ & 1.13$\times$ & 0.98$\times$ & 0.13$\times$ & 0.29$\times$ & 1$\times$ & 0.25$\times$ & 6.57$\times$ \\
         \hline
         Efficiency Improvement\^ & $^{\dag}$57$\times$ & 6.94$\times$ & 1.1$\times$ & 0.4$\times$ & 1$\times$ & 0.89$\times$ & 0.13$\times$\\
         \hline
    \end{tabular}
    
    \vspace{0.1cm}
    {\raggedright *Evaluations based on Llama-8B; \textasciicircum Nvidia H100 as Baseline; $^{\dag}$with CCPG\\}
    
    \label{tab:cross_platform}
\end{table*}

\subsection{System Scalability and Chiplet Clustering}
With the chiplet clustering scheme shown in Fig.~\ref{fig:chip_cluster}, only one cluster is fully activated while for all other clusters, only the scratchpad modules stay active for context window KV caching.
Fig.~\ref{fig:cluster_compare} shows the system power and energy efficiency improvements after implementing the chiplet clustering and power gating scheme (CCPG).
As much as 80\% power is saved for Llama-8B, leading to 57$\times$ energy efficiency improvement over Nvidia H100 with similar throughput. The results indicate that the larger the model size, \textit{i.e.} higher number of compute-tile chiplets, the greater the reduction of system power by CCPG. This is due to the higher ratio of chiplets that can be put to sleep mode. Under CCPG, the system power scales at $O(\log n$); thus, PICNIC is highly scalable to accommodate larger LLMs.

\begin{figure}[t]
    \centering
    \includegraphics[width=1\linewidth]{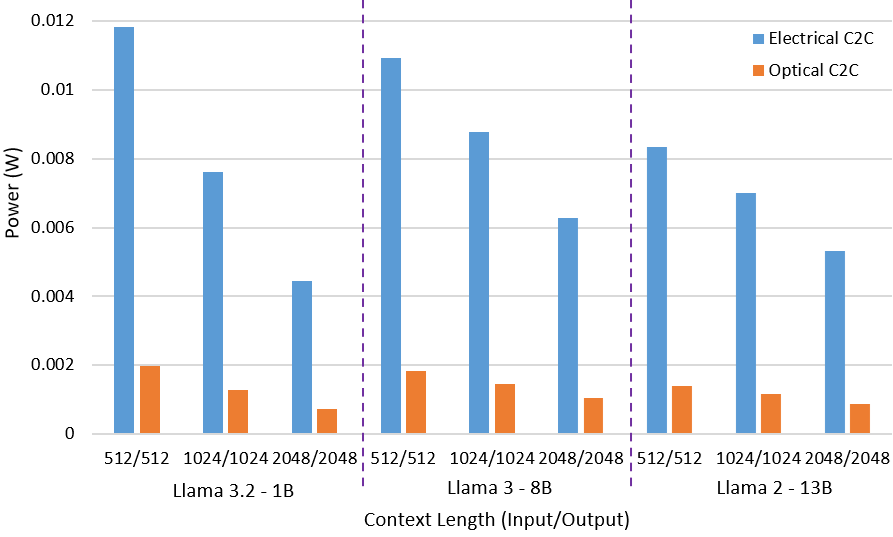}
    \caption{Average Power of C2C Data Transfer for Different Models and Context Lengths (Electrical vs Optical Interconnects)}
    \label{fig:c2c_power}
\end{figure}

\begin{figure}[t]
    \centering
    \includegraphics[width=0.85\linewidth]{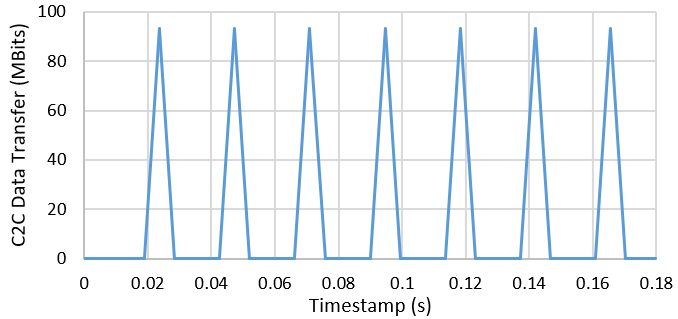}
    \caption{C2C Data Transfer Distribution Over Time (Llama 3.2-1B)}
    \label{fig:c2c_distrb}
\end{figure}

\subsection{Chip-to-chip (C2C) Communication}
The average power of C2C data transfer during LLM inference is affected by both model size and context length as illustrated in Fig.~\ref{fig:c2c_power}.
As the context length or model size increases, more computations are performed within the 2D-mesh within a chiplet, which cause higher computational latency, especially during the LLM decode phase.
Thus, the average C2C data transfer rate reduces because C2C communication occurs only after the computations in the 2D-mesh are completed.
As illustrated in Fig.~\ref{fig:c2c_distrb}, C2C data transfer occurs during certain time periods; apart from that, data movement and computations occur within IPCN and PEs of individual chiplet.

\subsection{Power and Area Breakdown}
The power and area breakdown of our PICNIC macro are shown in Table \ref{tab:pwr_and_area}.
Each pair of IPCN router-and-PE consumes 259 $uW$ of power at an area of 0.1842 $mm^2$.
The non-volatile RRAM PE consumes most of the area and power as it stores the model weights and also performs weighted-MAC (SMAC) within the same macro.
The power consumption of a router is comparable to that of the RRAM PE as it performs both data communication and in-network computing at the same time.

\begin{table}[t]
    \caption{Power \& Area Breakdown of PICNIC Macros (Unit)}
    \centering
    \setlength{\tabcolsep}{4.8pt}
    \begin{tabular}{c|cc|cc}
         \hline
         Macro & Power ($uW$) & Breakdown & Area ($mm^2$) & Breakdown \\
         \hline
         IMC PE \cite{rram_imc} & 120 & 46.3\% & 0.1442 & 78.3\% \\
         \hline
         Scratchpad & 42 & 16.2\% & 0.013 & 7.1\% \\
         \hline
         Router & 97 & 37.5\% & 0.025 & 13.5\% \\
         \hline
         TSVs & - & - & 0.002 & 1.1\% \\
         \hline \hline
         Total & \multirow{2}{*}{259} & \multirow{2}{*}{100\%} & \multirow{2}{*}{0.1842} & \multirow{2}{*}{100\%} \\
         (IPCN-PE) & & & & \\
         \hline \hline
         Softmax & 5.31 & - & 0.041 & - \\
         \hline
    \end{tabular}

    \vspace{0.1cm}
    {\raggedright \#Technology node: 7 nm $\vert$ Area per Compute Tile Chiplet: 189.6 $mm^2$ \\}
    \label{tab:pwr_and_area}
\end{table}

\section{Conclusion}
The PICNIC LLM inference accelerator with 3D IC chiplet design, consisting of Inter-PE Computational Network (IPCN) and non-volatile RRAM in-memory computing PE, interconnected via silicon photonics and coupled with efficient hardware scheduling scheme is shown to have superior performance and energy efficiency as compared to GPUs. It achieves 3.95$\times$ speedup and 30$\times$ efficiency improvement over the Nvidia A100 in Llama-8B inference (309 tokens/s and 10.9 tokens/J respectively) before CCPG. Furthermore, the implementation of CCPG on PICNIC reduces the system power further by 80\%, achieving 57$\times$ energy efficiency improvement over Nvidia H100 at similar throughput. As CCPG plays an increasingly pivotal role in scaling to larger models, PICNIC is a highly scalable architecture designed to operate within stringent power constraints. By ensuring sub-linear power scaling, PICNIC effectively mitigates communication bottlenecks through efficient compute-memory-network integration and compute resources orchestration.

\bibliographystyle{IEEEtran}
\bibliography{references}

\end{document}